\documentclass[letterpaper]{article} 
\usepackage{aaai2026}  
\usepackage{times}  
\usepackage{helvet}  
\usepackage{courier}  
\usepackage[hyphens]{url}  
\usepackage{graphicx} 
\usepackage{subcaption}
\usepackage{natbib}  
\usepackage{caption} 
\usepackage{algorithm}
\usepackage{algorithmic}
\usepackage{xcolor}
\usepackage{newfloat}
\usepackage{listings}
\usepackage{multirow}
\DeclareCaptionStyle{ruled}{labelfont=normalfont,labelsep=colon,strut=off} 
\floatstyle{ruled}
\newfloat{listing}{tb}{lst}{}
\floatname{listing}{Listing}
\usepackage{booktabs}

\newcommand{\qg}[0]{\textsc{Quotegraph}\xspace}
\newcommand{\qb}[0]{\textsc{Quotebank}\xspace}
\newcommand{\mycaption}[2]{\caption[#1]{\textbf{#1.} #2}}

\DeclareSymbolFont{extraup}{U}{zavm}{m}{n}
\DeclareMathSymbol{\zagreb}{\mathalpha}{extraup}{87}
\DeclareMathSymbol{\epfl}{\mathalpha}{extraup}{86}
\DeclareMathSymbol{\konstanz}{\mathalpha}{extraup}{84}
\DeclareMathSymbol{\au}{\mathalpha}{extraup}{81}

\title{Quotegraph: A Social Network Extracted from Millions of News Quotations}
\author {
    \textbf{Marko Čuljak}$^{\zagreb}$,
    \textbf{Robert West}$^{\epfl}$,
    \textbf{Andreas Spitz}$^{\konstanz}$,
    \textbf{Akhil Arora}$^{\au}$
}

\affiliations {
    $^\zagreb$University of Zagreb\quad 
    $^{\epfl}$EPFL\quad
    $^{\konstanz}$University of Konstanz\quad
    $^{\au}$Aarhus University \\
    marko.culjak@fer.unizg.hr
}

\usepackage{bibentry}

\usepackage{xspace}
\usepackage{amsmath}
\usepackage{amsfonts}
\usepackage{url}
\usepackage{marvosym}
\usepackage{ifthen}

\newcommand{\chatoDisplayMode}[1]{#1}

\newcommand{\inred}[1]{{\color{MyRed}\sf\textbf{\textsc{#1}}}}
\newcommand{\frameit}[2]{
  \begin{center}
  {\color{MyRed}
  \framebox[.9\columnwidth][l]{
    \begin{minipage}{.85\columnwidth}
    \inred{#1}: {\sf\color{MyBlack}#2}
    \end{minipage}
  }\\
  }
  \end{center}
}

\newcommand{\note}[2][]{\chatoDisplayMode{\def\@tmpsig{#1}\frameit{{\Pointinghand} Note}{#2\ifx \@tmpsig \@empty \else \mbox{ --\em #1}\fi}}}
\newcommand{\todo}[2][]{\chatoDisplayMode{\def\@tmpsig{#1}\frameit{{\Writinghand} To-do}{#2\ifx \@tmpsig \@empty \else \mbox{ --\em #1}\fi}}}

\newcommand{\abbrevStyle}[1]{#1}

\newcommand{\ie}{\abbrevStyle{i.e.}\xspace}
\newcommand{\eg}{\abbrevStyle{e.g.}\xspace}

\newcommand{\xhdr}[1]{\vspace{1.7mm}\noindent{{\bf #1.}}}

\newcommand{\textcite}[1]{\citeauthor{#1} \shortcite{#1}}

\newcommand{\hide}[1]{}

\makeatletter
\newcommand{\iffont}[2]{\ifthenelse{\equal{\f@family}{#1}}{#2}{}}
\makeatother

\iffont{ptm}{
  \usepackage{mathptmx}

  \DeclareSymbolFont{greek}{OML}{cmm}{m}{n}
  \DeclareMathSymbol{\alpha}{\mathalpha}{greek}{"0B}
  \DeclareMathSymbol{\beta}{\mathalpha}{greek}{"0C}
  \DeclareMathSymbol{\gamma}{\mathalpha}{greek}{"0D}
  \DeclareMathSymbol{\delta}{\mathalpha}{greek}{"0E}
  \DeclareMathSymbol{\epsilon}{\mathalpha}{greek}{"0F}
  \DeclareMathSymbol{\zeta}{\mathalpha}{greek}{"10}
  \DeclareMathSymbol{\eta}{\mathalpha}{greek}{"11}
  \DeclareMathSymbol{\theta}{\mathalpha}{greek}{"12}
  \DeclareMathSymbol{\iota}{\mathalpha}{greek}{"13}
  \DeclareMathSymbol{\kappa}{\mathalpha}{greek}{"14}
  \DeclareMathSymbol{\lambda}{\mathalpha}{greek}{"15}
  \DeclareMathSymbol{\mu}{\mathalpha}{greek}{"16}
  \DeclareMathSymbol{\nu}{\mathalpha}{greek}{"17}
  \DeclareMathSymbol{\xi}{\mathalpha}{greek}{"18}
  \DeclareMathSymbol{\pi}{\mathalpha}{greek}{"19}
  \DeclareMathSymbol{\rho}{\mathalpha}{greek}{"1A}
  \DeclareMathSymbol{\sigma}{\mathalpha}{greek}{"1B}
  \DeclareMathSymbol{\tau}{\mathalpha}{greek}{"1C}
  \DeclareMathSymbol{\upsilon}{\mathalpha}{greek}{"1D}
  \DeclareMathSymbol{\phi}{\mathalpha}{greek}{"1E}
  \DeclareMathSymbol{\chi}{\mathalpha}{greek}{"1F}
  \DeclareMathSymbol{\psi}{\mathalpha}{greek}{"20}
  \DeclareMathSymbol{\omega}{\mathalpha}{greek}{"21}
  \DeclareMathSymbol{\varepsilon}{\mathalpha}{greek}{"22}
  \DeclareMathSymbol{\vartheta}{\mathalpha}{greek}{"23}
  \DeclareMathSymbol{\varpi}{\mathalpha}{greek}{"24}
  \DeclareMathSymbol{\varrho}{\mathalpha}{greek}{"25}
  \DeclareMathSymbol{\varsigma}{\mathalpha}{greek}{"26}
  \DeclareMathSymbol{\varphi}{\mathalpha}{greek}{"27}
  \DeclareSymbolFont{otone}{OT1}{cmr}{m}{n}
  \DeclareMathSymbol{\Gamma}{\mathalpha}{otone}{0}
  \DeclareMathSymbol{\Delta}{\mathalpha}{otone}{1}
  \DeclareMathSymbol{\Theta}{\mathalpha}{otone}{2}
  \DeclareMathSymbol{\Lambda}{\mathalpha}{otone}{3}
  \DeclareMathSymbol{\Xi}{\mathalpha}{otone}{4}
  \DeclareMathSymbol{\Pi}{\mathalpha}{otone}{5}
  \DeclareMathSymbol{\Sigma}{\mathalpha}{otone}{6}
  \DeclareMathSymbol{\Upsilon}{\mathalpha}{otone}{7}
  \DeclareMathSymbol{\Phi}{\mathalpha}{otone}{8}
  \DeclareMathSymbol{\Psi}{\mathalpha}{otone}{9}
  \DeclareMathSymbol{\Omega}{\mathalpha}{otone}{10}
  \DeclareSymbolFont{syms}{OML}{cmm}{m}{it}
  \DeclareMathSymbol{\partial}{\mathord}{syms}{"40}
  \DeclareMathAlphabet{\mathbold}{OML}{cmm}{b}{it}
  \DeclareSymbolFont{largesymbols}{OMX}{cmex}{m}{n}

}

\begin{document}

\maketitle

\begin{abstract}
We introduce \qg, a novel large-scale social network derived from speaker-attributed quotations in English news articles published between 2008 and 2020. \qg consists of 528 thousand unique nodes and 8.63 million directed edges, pointing from speakers to persons they mention. The nodes are linked to their corresponding items in Wikidata, thereby endowing the dataset with detailed biographic entity information, including nationality, gender, and political affiliation. Being derived from \textsc{Quotebank}, a massive corpus of quotations, relations in \qg are additionally enriched with the information about the context in which they are featured. Each part of the network construction pipeline is language agnostic, enabling the construction of similar datasets based on non-English news corpora. We believe \qg is a compelling resource for computational social scientists, complementary to online social networks, with the potential to yield novel insights into the behavior of public figures and how it is captured in the news.

\end{abstract}

\begin{links}
    \link{Code}{https://github.com/au-clan/quotegraph}
    \link{Data}{https://zenodo.org/records/16275215}
\end{links}

\section{Introduction}
Understanding human interactions and characterizing the dynamics observed in social networks have been central themes of social science since its inception. Pioneering work in social network analysis is attributed to Jacob L. Moreno and his book \textit{Who shall survive?} \cite{moreno1934who}, in which he analyzed friendships in a social network of 33 schoolchildren. Nine decades later, the proliferation of data and advances in data processing have enabled the collection and analysis of social networks several orders of magnitude larger than before, such as Facebook, with 3.4 billion daily active users as of March 2025 \cite{meta2025meta}.

Despite significantly accelerating social science by providing a plethora of valuable insights into human behavior, modern online social networks (OSNs) are not without limitations. Platform constraints, content moderation, and easily deletable and modifiable posts can lead to unfaithful representations of users' opinions and interactions, potentially hiding their true opinions and intentions. In addition, the widespread use of powerful transformer-based language models has led to the web being plagued with misleading AI-generated content. To alleviate those limitations, we can benefit from the replication of the analyses conducted on OSNs on social networks from other sources.
\begin{figure}[t]
    \centering
    \includegraphics[width=\linewidth]{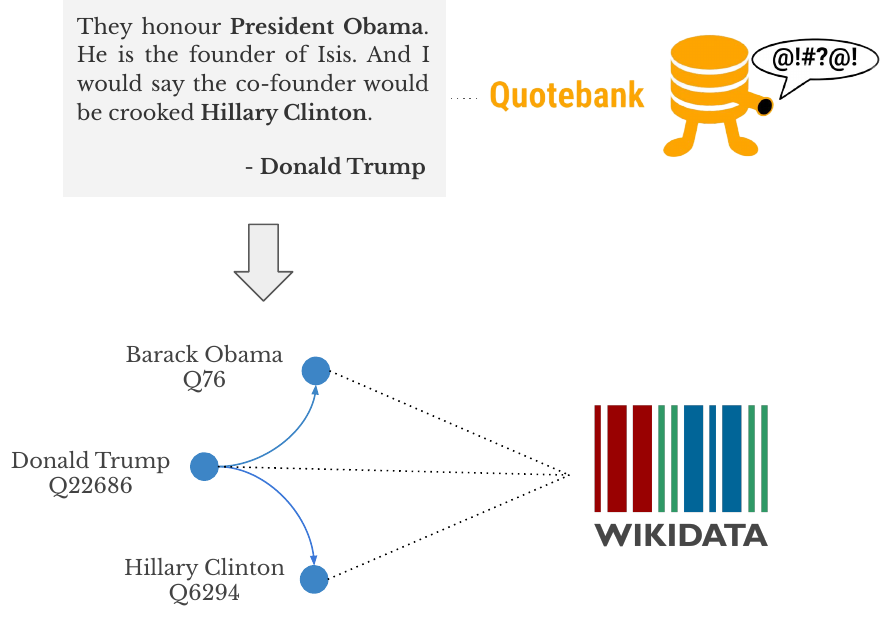}
    \caption{\textbf{\qg building illustration.} In this example, based on the quotation from Quotebank in the upper left corner, we draw a directed edge from the speaker of the quotation, Donald Trump (Q22686), to Barack Obama (Q76) and Hillary Clinton (Q6294), whom he mentioned in his quotation. All three nodes have a corresponding item in Wikidata identified uniquely by its QID.}
    \label{fig:network-construction}
\end{figure}

However, explicit social networks of comparable substance and volume are impossible to construct outside of online platforms. Therefore, we turn to implicit social networks that can be extracted from abundant textual corpora \cite{elson2010extracting, deleris2018know, bro2025frustratingly} as a resource complementary to OSNs. More specifically, we introduce \qg, a large-scale social network from \textsc{Quotebank} \cite{vaucher2021quotebank}, a massive corpus of speaker-attributed news quotations by drawing directed edges between the speaker of the quotation and persons mentioned in it. We illustrate the network construction process in Fig. \ref{fig:network-construction}. By constructing \qg, we continue previous work on facilitating quotation analysis \cite{pavllo2018quootstrap, vaucher2021quotebank, vukovic2022quote, kulz2023united, hu2023quotativesb} and provide a novel large-scale resource for investigations into the behavior of public figures captured in the news. \qg boasts 528k nodes representing actors in public discourse linked to their respective Wikidata items \cite{vrandecic2014wikidataa} and 8.63M edges representing quotations linked to news articles in which they appear. 

\section{Preliminaries}
\xhdr{\qb preliminaries} \qb has been released in the two formats: (1) quotation-centric and (2) article-centric format. The former is a set of all extracted unique quotations with the name of the most likely speaker and the accompanying metadata, including the URLs of all the articles in which it appears, the earliest appearance date, and, for each of the speaker candidates, the probability that they uttered the quotation. The quotation-level data, however, does not contain the context of the articles from which they are extracted. This field is available as part of the article-centric format, along with article metadata, quotation offsets, spans of named entity mentions, and local speaker attribution probabilities. Those local probabilities are aggregated to obtain global attribution probabilities in the quotation-centric data. The contents of the articles are tokenized and stored as lists of Stanford CoreNLP tokens \cite{manning2014stanford}. For a more detailed description of the \qb formats, please check \qb's Zenodo record \cite{vaucher2021quotebankzenodo}.

\xhdr{Network building prerequisites} As a prerequisite for building \qg, we first require a dataset of speaker-attributed quotations with the accompanying context, with the extracted person-type entity spans. Second, we must link the mentions of the actors and their Wikidata items. Since the first prerequisite is already satisfied in the original \qb corpus, we must satisfy the second prerequisite, which boils down to solving the entity disambiguation task. Although state-of-the-art methods perform well on the task, they are infeasible to run at \qb's scale. To disambiguate all the 573M entity mentions in \qb using ReFinED \cite{ayoola2022refineda}, the most resource-efficient method, it would take approximately 95 days to complete the task on an academic budget. Therefore, we use lightweight heuristics proposed by \citet{culjak2022strong} implemented as a part of \texttt{quotebank-toolkit}\footnote{https://github.com/epfl-dlab/quotebank-toolkit}\cite{culjak2023auxiliary} and link all the entity mentions with Wikidata in approximately two hours. Additionally, since the heuristics require no training data, they are easily transferable to corpora in other languages, thereby making the entire network construction pipeline language-agnostic.

\section{Data preprocessing}
Having satisfied the prerequisites, we proceed to network construction. Although straightforward, \qg building is complicated by the noise present in Quotebank, its heterogeneity and scale. Thus, we perform the following preprocessing steps to improve the quality of the resulting network.

\xhdr{Short quotation removal} We observed that short quotations are either generic quotes that can be correctly attributed to multiple people (\eg, \textit{I love you.}) or only contain a person's name, a movie title, or a book title. Additionally, in some cases, the same word is repeated multiple times within a single quotation (\eg, \textit{Trump, Trump, Trump, Trump, Trump!}). Thus, we discard all quotations with less than $l_q$ case-folded unique words, not considering punctuation, and use $l_q = 5$.

\xhdr{Grouping similar quotations} A quotation can appear in slightly different forms or as a part of a longer quotation in various news outlets \cite{pavllo2018quootstrap}. Although a high frequency of a quote in different articles could signify relevance, due to the noise and heterogeneity of \qb, it could also reduce the network quality. Furthermore, different versions of the same quotation can be attributed to different speakers, thus leading to the emergence of non-existent interactions. Therefore, following \citet{pavllo2018quootstrap}, we group all the quotations that, when case folded, share the same substring of length at least $l_s = 8$ words excluding punctuation, and substitute them with the longest quotation in the group.

\xhdr{Discarding spurious mentions} Some entity names contain punctuation and stopwords. If any entity alias contains such tokens,  each of their occurrences in an article will be identified as a mention of the entity due to the limitations of the system for speaker candidate extraction \cite{pavllo2018quootstrap}. To resolve this issue, we do not consider (1) one-character tokens, (2) tokens with no alphabetical characters, and (3) stopwords as possible mentions for edge extraction.

\xhdr{Determining quotation spans} Since neither the quotation length in tokens nor the position of their ending tokens is provided in the article-centric \qb, we must find ending tokens for each quotation because it will be necessary for the identification of entity mentions that appear inside quotations. Owing to the use of Stanford CoreNLP \cite{manning2014stanford} tokenizer for the initial tokenization of \qb \cite{vaucher2021quotebank}, finding the positions of the ending tokens is trivial since the tokenizer conveniently distinguishes between opening and closing quotation marks. Thus, to determine the position of the last token in a quotation, we iterate through all the tokens in an article, starting with the starting token of the quotation, until we reach the closing quotation mark. Following \textcite{vaucher2021quotebank}, who extract the innermost quote in the case of nested quotes, we assume that the first closing quotation mark found does not mark the ending of any inner quotation. 

\xhdr{Finding mentions appearing in quotations} Having defined the ending token for each quotation, the mentions appearing in quotations could be identified by comparing their token spans with the quotation spans. However, we found that the quotation spans and name entity mention spans are not always consistent, which can cause the naive approach to yield incorrect results at times. To this end, we devise a simple approach based on string matching. We first extract all the tokens of all the quotations in an article and concatenate them via spaces, obtaining a string $s_q$. We repeat the same procedure for the tokens of the named entity mentions based on the spans listed in article-centric \qb, and obtain a string $s_m$ for each mention. To determine whether a mention appears inside a quotation, we check if $s_m$ is a substring of $s_q$.

\section{Network construction and postprocessing}
After preprocessing, edges between the actors can be trivially constructed by connecting each speaker with the actors they mention in their quotations. An edge in \qg is uniquely identified by the (speaker Wikidata QID, mention Wikidata QID, quote identifier) triplet. To further improve the dataset quality, we perform the following postprocessing steps.

\xhdr{Aggregating speakers over contexts} To attribute each quotation to the globally most probable speaker QID, we first set the local attribution probabilities of the entities to the previously computed local attribution probabilities of their respective names. Following \citet{vaucher2021quotebank}, we sum the local probabilities over all the quotation contexts and attribute each quotation to the entity with the highest global attribution probability.

\xhdr{Aggregating mentions over contexts} We aggregate the mentions over contexts by selecting the most common mention as the target node of an edge in \qg. If a quotation contains multiple mentions, we choose the most common set of Wikidata entities within the quote as the target nodes and create a distinct edge for each mentioned entity in the obtained set.

\xhdr{Removing self-loops} Due to the limitations of quotation attribution in Quotebank, some quotations are erroneously attributed to the same persons mentioned in them. Thus, the edges derived from such quotations are self-loops.

\section{Processing Wikidata node attributes}
\begin{table}[t]
    \mycaption{Top-level occupations used to define respective domains in the left column}{}
    \centering
    \begin{tabular}{cc}
    \toprule
        \textbf{Domain} & \textbf{Top-level occupation}\\
        \midrule
        Art & artist (Q483501), creator (Q2500638)\\
        Politics & politician (Q82955), lawyer (Q185351)\\
        Sport & sportsman (Q50995749)\\
    \bottomrule
    \end{tabular}
    \label{tab:occupations}
\end{table}

\begin{table}[h!]
\centering
\mycaption{Structural graph property metrics and their values for \qg}{The mean degree is computed as the total degree, i.e., by taking into account both indegree and outdegree. CC stands for connected
components, more specifically, weakly connected components.}
    \begin{tabular}{lr}
        \toprule
        \textbf{Metric} & \textbf{Value} \\
        \midrule
        Nodes & 528k \\
        Edges & 8.63M \\
        Mean degree & 32.77 \\
        \#CC & 5651 \\
        \% Nodes in the largest CC & 97.53 \\
        Degree assortativity & 0.034 \\
        Global clustering coefficient & 0.265 \\
        \midrule
        Assortative mixing by: & \\ 
        \quad\quad Nationality & 0.326 \\
        \quad\quad Occupation & 0.283 \\
        \quad\quad Gender & 0.052 \\
        \quad\quad Party: & \\ 
        \qquad\qquad USA & 0.024 \\
        \qquad\qquad UK & 0.121 \\
        \qquad\qquad India & 0.072 \\
        \bottomrule
    \end{tabular}
    \label{tab:structural-properties}
\end{table}

In addition to precise identification of the actors, the link with Wikidata enables access to entity information stored in the form of (property, value) pairs. Being crowdsourced, the way information is presented in Wikidata is not always consistent. Therefore, below we detail selected properties and the preprocessing steps to facilitate the utilization of Wikidata information.

\xhdr{Date of birth (P569)} While a person may have multiple dates of birth listed as a part of their Wikidata item, this is the case for only 0.1\% of entities in all the extracted networks. Thus, in those cases, we extract the first listed birth date.

\xhdr{Nationality (P27)} A person can have multiple nationalities, so we extract all the nationalities listed and do not consider countries that no longer exist, such as the Socialist Federal Republic of Yugoslavia (Q838261) or the Soviet Union (Q15180).

\xhdr{Gender (P21)} For gender, we consider three categories: female, male, and other. We deem an entity to be female or male if only female (Q6581072) or male gender (Q6581097) is listed in their gender statement. If either a non-binary gender (Q48270) or multiple genders are listed, we label the gender as 'other'.

\xhdr{Political party affiliation (P102)} A person can switch their political party throughout their life. In this case, we assign a party to an entity based on the date of the quotation and Wikidata information about the start and end dates of the party affiliation. If this information is not provided, we select the last party listed as it is likely the most recent one. Suppose only a year or a month of the start or the end of the party affiliation is provided. In that case, we extract the party an entity is affiliated with at the earliest possible date that meets the provided information. For example, if a person switched their party in 2008, we assume that the switch happened on January 1st, 2008.

\xhdr{Occupation (P106)} We split the occupations into domains based on the Wikidata occupation hierarchy modeled by the \textit{subclass of} property (P279). We define one or more top-level occupations for each domain and consider all the occupations below them in the hierarchy to belong to the specified domain. We focus on art, sport, and law \& politics domains and label other occupations as other. We present the domain-to-top-level occupation mapping in Tab. \ref{tab:occupations}. Since art can take many forms and can be performed on various levels of commitment, we deem an entity an artist if their domain is solely art, rather than politics or sports. For example, according to Wikidata, Donald Trump (Q22686) is considered an actor (Q33999) and a writer (Q36180), both of which belong to the artist (Q483501) occupation tree. We do not disentangle other overlapping domains.

\begin{figure*}[ht]
    \centering
    \subfloat[\centering Indegree distribution]{{\includegraphics[width=.33\textwidth]{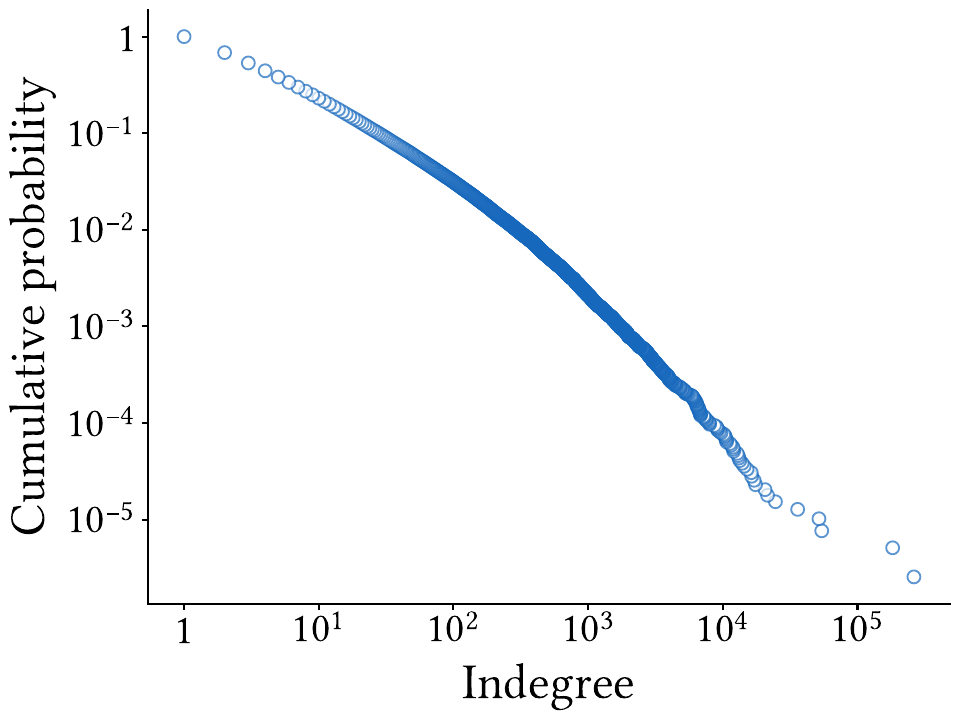}}\label{fig:indeg-dist}}
    \subfloat[\centering Nationality distribution]{{\includegraphics[width=.33\textwidth]{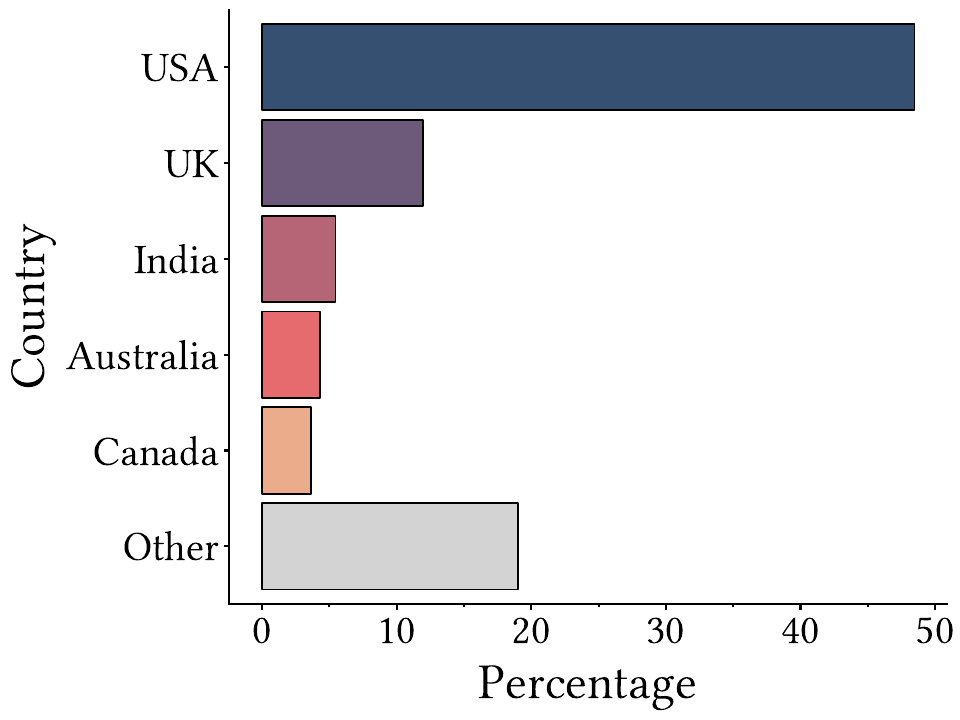}}\label{fig:nation-dist}}%
    \subfloat[\centering Age distribution]{{\includegraphics[width=.33\textwidth]{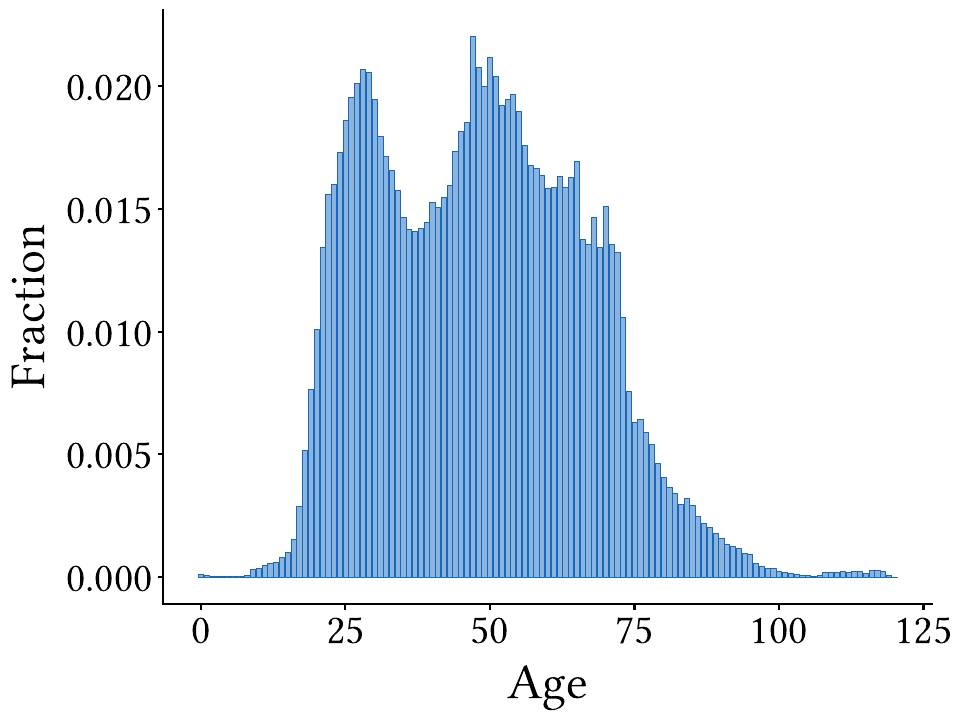}}\label{fig:age-dist}}\\
    \subfloat[\centering Outdegree distribution]{{\includegraphics[width=.33\textwidth]{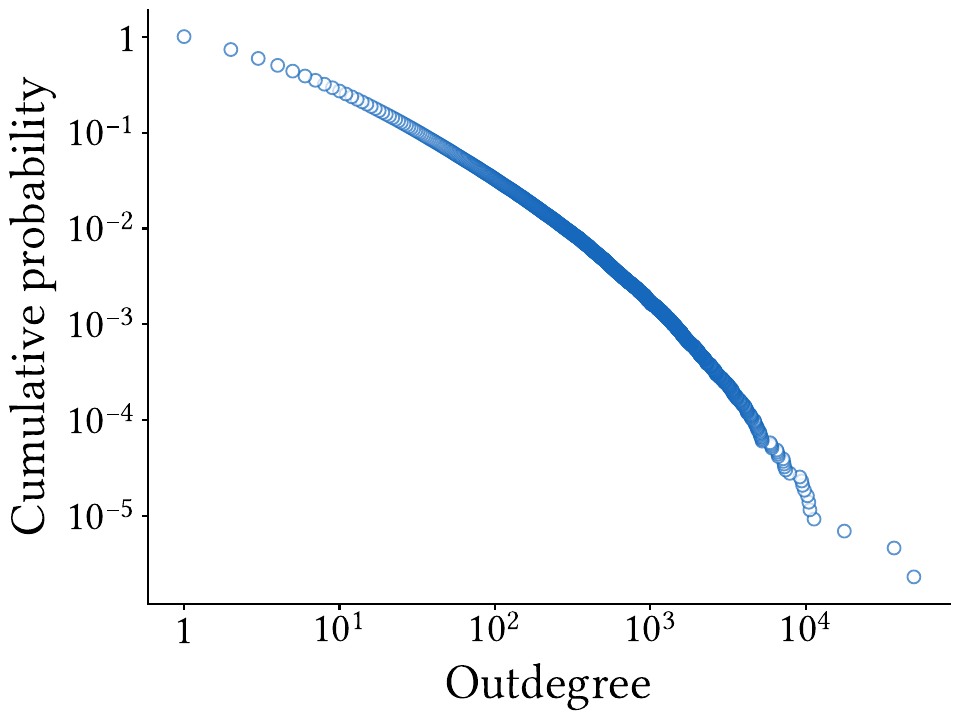}}\label{fig:outdeg-dist}}
    \subfloat[\centering Occupation distribution]{{\includegraphics[width=.33\textwidth]{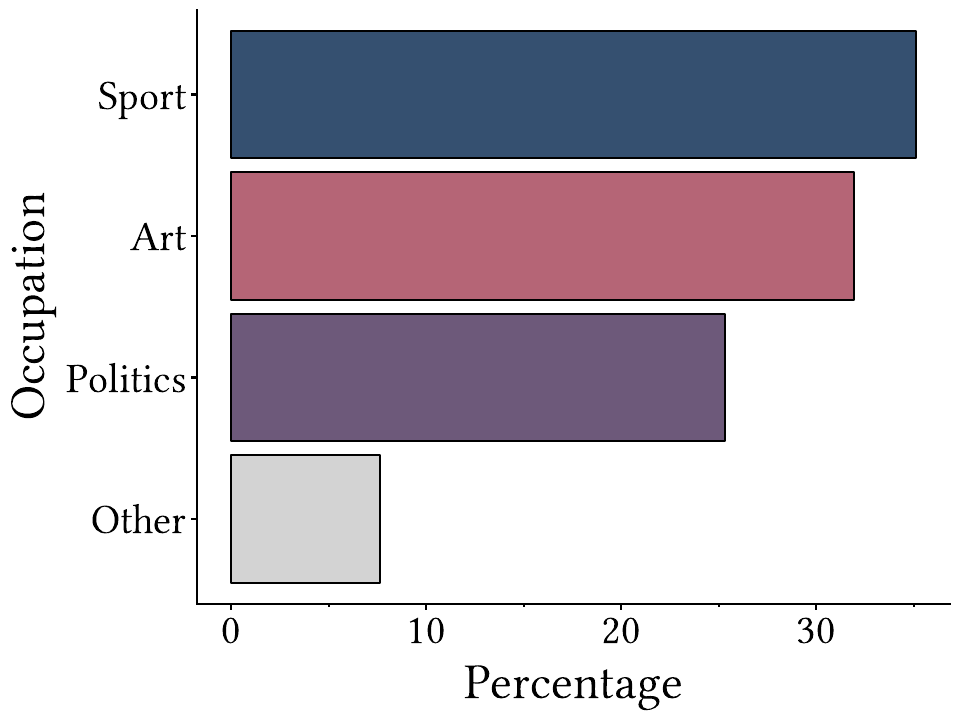}}\label{fig:occupation-dist}}
    \subfloat[\centering 10 most central nodes]{{\includegraphics[width=.33\textwidth]{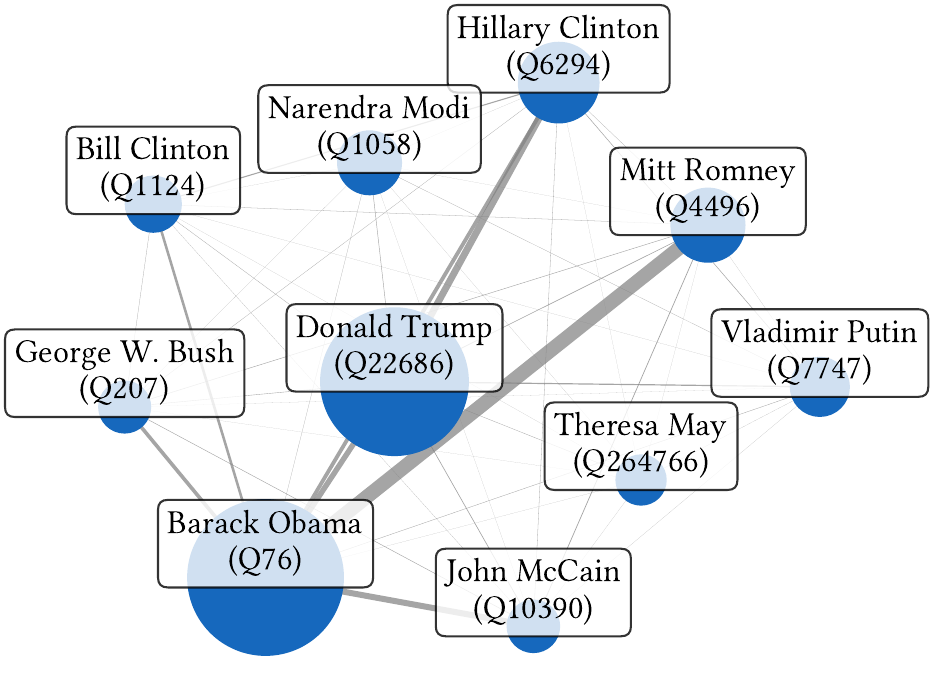}}\label{fig:central-nodes}}
    \mycaption{An overview of \qg}{Fig. \ref{fig:indeg-dist} and \ref{fig:outdeg-dist} show the cumulative indegree and outdegree distributions. On the y-axis, we show the probability $P(d \geq d_k)$ for a degree $d_k$ on the x-axis. Fig. \ref{fig:nation-dist}, \ref{fig:occupation-dist}, and \ref{fig:age-dist} present the distributions of nationality, occupation, and age, respectively. Fig. \ref{fig:central-nodes} is a visualization of a subnetwork consisting of the 10 most central nodes estimated by PageRank. Node size is proportional to the total degree of a node, while edge width is proportional to the number of unique interactions between the actors in both directions.}%
    \label{fig:overview}%
    \vspace{-0.5em}
\end{figure*}

\subsection{Structural properties}

In Table \ref{tab:structural-properties}, we present an overview of the metrics that describe the structural properties of \qg. The network is predominantly organized as a single connected component with 97.5\% of all the nodes contained within. Although this phenomenon is common in real-world social networks \cite{davis1997corporate,amaral2000classes,newman2004coauthorship}, in \qg, it is further amplified by the interactions that are artifacts of imperfect speaker attribution and entity disambiguation.
\qg has positive degree assortativity, heavy-tailed degree distributions (Fig. \ref{fig:indeg-dist} and \ref{fig:outdeg-dist}), and a high clustering coefficient. All of the listed properties are typical for large-scale real-world social networks. 

\xhdr{Assortative mixing} Using the definition of modularity proposed by \textcite{newman2003mixing}, we investigate the existence of assortative mixing by nationality, domain, gender, and party affiliation. \qg shows high assortative mixing by nationality and domain, and much lower assortative mixing by gender and party affiliation. We also observe that the UK is characterized by relatively high modularity by party compared to the USA and India. However, understanding the causes of this phenomenon is beyond the scope of this work and warrants further study, which we leave for future work.

\subsection{Demographics of \qg}
Next, we provide an overview of the demographics of actors in \qg. We calculate the distributions of node features based on edge ends rather than individual nodes. In other words, our feature distributions are degree-weighted, meaning each node's features contribute to the overall distribution in proportion to its total degree. The fractions in the computed distributions model the probability that a randomly chosen end of a random edge is attached to a node with a particular feature.

\xhdr{Nationality} In Fig. \ref{fig:nation-dist}, we show the distribution of the nationalities in the networks. Overall, actors from the USA are prevalent (48.4\%). Still, there is a non-negligible number of interactions with actors from other nationalities, albeit predominantly English-speaking.

\xhdr{Occupation} As shown in Fig. \ref{fig:occupation-dist}, sportsmen are the most represented occupation category in \qg. However, the most central nodes according to PageRank \cite{page1999pagerank} are politicians (Fig. \ref{fig:central-nodes}).

\xhdr{Age} The domain distributions are also reflected in the respective age distributions (\ref{fig:age-dist}). We observe two peaks, one around the age of 28, explainable by the prevalence of sportspeople  \qg (Fig. \ref{fig:occupation-dist}), and one after the age of 47, likely due to the high proportion of interactions between politicians.

\xhdr{Gender} In \qg, men appear in approximately 87\% of interactions, highlighting a strong gender bias in news coverage.

\section{Example application: nominal gender bias}
To highlight the potential of \qg for computational social science research, we utilize it to reproduce the findings of \textcite{atir2018how}, who studied gender bias in the use of first vs. last name when referring to professionals. Focusing on how the general public perceives professionals, Atir and Ferguson found that women are significantly more likely to be referred to by their first name rather than their last name, which causes them to be perceived as less prominent. Additionally, \textcite{marjanovic2022quantifying} observed a similar nominal gender bias on Reddit.
 
 We utilize \qg's scale and its link to Wikidata to control for biographical features of the actors (occupation, nationality, and age), their prominence (approximated via PageRank), and features related to their names (length and frequency) and model how gender of a person mentioned in a quotation is related to whether they are referred to by their first, last or full name. Having obtained findings consistent with the aforementioned studies \ie women being approximately twice as likely to be referred to by their first name than men, we found that this nominal gender bias is also present in communication between professionals and not limited to communication in the general public. We leave further exploration of the causes underpinning the reference choice for future work.

\section{Related work}
\xhdr{Quotation attribution and analysis}
Automatic quotation attribution, \ie mapping quotations to their respective speakers, has been studied in literary texts \cite{elson2010automatic, muzny2017twostage, vishnubhotla2023improving} and news articles \cite{okeefe2012sequence, pareti2013automatically, almeida2014joint, newell2018quote, pavllo2018quootstrap, vaucher2021quotebank, zhang2022directquote}. Being the largest available quotation corpus, \qb is the basis of several works in quotation analysis with \textcite{kulz2023united} analyzing the increase in negativity in quotes by US politicians following the 2016 Primary Campaigns, \textcite{vukovic2022quote} building an interface for exploration of \qb, and \textcite{hu2023quotativesb} studying the increasing use of non-objective quotatives as evidence for the gradually decreasing objectivity in the news. In addition to curation and analysis of automatically constructed speaker-attributed quotation corpora, recent works also explore crowdsourced corpora such as Wikiquote \cite{giammona2019print, kuculo2022quotekg} or manually annotated Common Crawl\footnote{https://commoncrawl.org/} \cite{palowitch2025socialquotes}.

\xhdr{Extraction and analysis of implicit social networks} The extraction and analysis from unstructured data has been extensively studied. Social networks have been extracted from various types of corpora such as emails \cite{culotta2004extracting, diesner2005communication, eckmann2004entropy, tyler2005mail, michalski2011matching, dinh2023enhancing}, scientific citations \cite{lehmann2003citation, price1965networks, touwen2024learning, vsubelj2016clustering, sen2008collective} and collaborations \cite{barabasi2002evolution, moody2004structure, newman2004coauthorship, Newman2001ScientificCNA, Newman2000ScientificCNA, Newman2000TheSOA, Fronczak2021ScientificSFA}, and Wikipedia \cite{geiss2015friendships}. Other works also focus on extracting social networks of characters from literary works \cite{alice, validating, fiction} and historical records \cite{medici, bacon}. The most closely related works to \qg are those that aim to extract social networks from news corpora. Recently, \textcite{bro2025frustratingly} used GPT-4 to extract a social network from Chinese news. In addition, \textcite{pouliquen2008extracting} built a social network on multilingual news using co-occurrence statistics, and quotation-based relationships similar to \qg. While valuable, the aforementioned two news-based social networks are several orders of magnitude smaller in scale than \qg. To our knowledge, \qg is the largest publicly available social network of this type. 

\section{Potential applications and conclusions}
We believe that \qg's scale and richness, spanning more than a decade, make it a compelling playground for computational social scientists. First, \qg can be used to study the evolution of political polarization among political actors and its relation to growing political polarization among the general public \cite{pew2014polarization}. The dataset can also be used to explore the properties of the actors that frequently mention other actors and identify periods in which political debate is focused more on persons than political issues. Second, the network structure of \qg can be used to study the emergence and properties of communities that transcend already defined groups, such as political parties, potentially revealing cross-partisan alliances. Third, owing to its integration with Wikidata, the dataset enables large-scale analyses of biases revolving around biographic features such as race and gender in public discourse. Additionally, one can study when and why politicians mention persons from other occupations and vice-versa. Lastly, each part of the network extraction pipeline is language-agnostic, making it applicable to news corpora in languages other than English.
\newpage

\bibliography{aaai2026}

\end{document}